\documentclass[amsmath,amssymb,twocolumn,aps,notitlepage,showpacs,pre,floatfix,longbibliography]{revtex4-1}
\usepackage{graphicx}% Include figure files
\usepackage{bm}% bold math
\usepackage{hyperref}% add hypertext capabilities
\usepackage[usenames, dvipsnames]{color}
\usepackage{tabto,lipsum} % add margin symbols
\usepackage{floatrow} % add side captions
\usepackage{amsmath}
\usepackage{csquotes}
\usepackage{float}

\hypersetup{
    colorlinks=true,
    linkcolor=blue,
    citecolor=red,
    filecolor=blue,
    urlcolor=blue,
}
\usepackage[normalem]{ulem}

\begin{document}

%\title{Out-of-equilibrium features of a non-reciprocal Brownian dimer}
\title{Non-reciprocity drives a Brownian dimer out of equilibrium}
%\date{\today}
%\pacs{05.20.-y, 04.40.-b, 05.90.+m}

\author{Suman Pramanik$^{1,3}$, Soham Dutta$^{1}$, and Arnab Saha$^{1,2}$}
\email{sahaarn@gmail.com}
\affiliation{$^1$  University of Calcutta, 92 A.P.C. Road, Kolkata, Pin-700009, India}
\affiliation{$^2$ Laboratoire de Physique Théorique et Modélisation, UMR 8089, CY Cergy Paris Université, 95302 Cergy-Pontoise, France}
\affiliation{$^3$ Gour Mohan Sachin Mandal Mahavidyalaya,P.O.- Bireswarpur, South 24 Parganas, West Bengal, Pin-743336, India}

\begin{abstract}
We consider the minimal model of a two-dimensional Brownian dimer consisting of two overdamped monomers, trapped in an isotropic harmonic potential and mutually coupled by a \emph{non-reciprocal} harmonic spring that violates Newton's action-reaction principle. We have shown that the non-reciprocal interaction alone can drive the system far from equilibrium, in the absence of any external time dependent drive and being in contact with a single thermal bath. The exact steady-state probability distribution and current are explicitly calculated for the zero-rest-length limit of the spring, which eventually maps our model to another non-equilibrium phenomenon, called \emph{Brownian gyration}. For a spring with finite rest length, these quantities are calculated numerically.
\end{abstract}

\maketitle

\section{Introduction}
The third law of Newtonian mechanics states that: \emph{`Actioni contrariam semper and æqualem esse reactionem: sive corporum duorum actiones in se mutuo semper esse æqualeset in partes contrarias dirigi.'} (that is, \emph{`To every action there is always opposed an equal reaction; or the mutual actions of two bodies upon each other are always equal, and directed to contrary parts.'}) \cite{chandrasekhar2003newton}. According to the pedagogical texts of classical mechanics in the context of inter-particle forces of a system of particles, this law leads to the mutual cancellation of all the inter-particle forces and conservation of momentum in the absence of any external drive. Thus, the mechanical stability for the system is established \cite{goldstein1950classical}. However, contemporary research on systems where the principle of momentum conservation and mechanical equilibrium are not applicable, has become significant. In the domain of classical physics, prominent examples of such systems are from the world of living or active matter \cite{marchetti2013hydrodynamics,bechinger2016active}, where the individual constituents consume energy from the surrounding as \enquote{fuel} to self-propel. Such entities or `particles' can be as small as a living cell in a tissue or can be as big as a bird in a flock or sheep in a herd \cite{barberis2016large, gomez2022intermittent}. Often, these particles interact with each other physically, chemically and/or socially, forming a group which neither cares for momentum conservation nor mechanical stability. Therefore, the inter-particle interactions in such a group are {\it not} necessarily reciprocal to maintain the action-reaction principle \cite{fruchart2021non}.

%%%%%%%%%%%%%%%%%%%%%%%%%%%%%%%%%%%%%%%%%%%%%%%%%%%%%%
Having said that, it is possible for an active particle to interact with others non-reciprocally --- here, we ask that whether or not the non-reciprocal inter-particle interaction (among a system of particles) can itself generate directed motion or produce non-zero currents within the system, driving it far from equilibrium. We will illustrate this with an analytically-tractable model consisting of a harmonically-confined Brownian dimer, where the constituent (apparently passive) monomers interact among themselves with a non-reciprocal harmonic spring. Though the constituent monomers are not moving by themselves, here we will show that the dimer can indeed produce a non-zero current, as observed in {\it microscopic gyration} \cite{dotsenko2022cooperative}, driving the system to a non-equilibrium steady-state (NESS) solely due to the non-reciprocal feature of the inter-monomer interaction.

%%%%%%%%%%%%%%%%%%%%%%%%%%%%%%%%%%%%%%%%%%%%%%%%%%%%%%%%%%

When the degrees of freedom (DoF) of a Brownian system interact with each other {\it{reciprocally}}, it is essential for the DoF to be in simultaneous contact with at least two thermal reservoirs at different temperatures, for the microscopic gyration of the system \cite{filliger2007brownian,mancois2018two,dutta2026microscopic}.  These DoF can be dynamically coupled via several ways, e.g. by an external anisotropic trap \cite{mancois2018two}, tensorial mobility of the particle due to its internal shape anisotropy \cite{dutta2026microscopic}, odd transport processes \cite{abdoli2026odd}, etc.  The heat that flows from one bath to the other throughout the DoF of the system (via these couplings), exerts a thermal torque on the system, such that it can gyrate and eventually drive the system towards a NESS, characterized by a non-zero gyration current \cite{chiang2017electrical,abdoli2026quadrupolar,bae2021inertial,cerasoli2018asymmetry,das2022inferring,chang2020autonomous}. In such systems, if the temperatures of the reservoirs become equal, there will be no heat flux, and consequently, the thermal torque along with the microscopic gyration vanishes --- hence, the system equilibrates. 

%Microscopic gyration occurs in Brownian systems when multiple degrees of freedom involved in the dynamics of a system interact with each other {\it{reciprocally}} and it is absolutely essential for the degrees of freedom to be in contact with different thermal reservoirs at different temperatures simultaneously. 

%%%%%%%%%%%%%%%%%%%%%%%%%%%%%%%%%%%%%%%%%%%%%%%%%%%%%%%%%

However, here we will employ a Brownian dimer to show that if the degrees of freedom interact with each other {\it{non-reciprocally}},  the aforesaid temperature gradient (or, the difference in temperatures of the thermal baths attached to DoF) {\it{does not}} remain an essential ingredient to obtain a finite current, as was indispensable for the microscopic gyration of a Brownian system. The unbalanced force, generated by the non-reciprocal interaction among the particles in contact with a single bath, is enough for the system to generate the current. This autonomous gyration eventually drives the system to a NESS (as in the earlier cases of microscopic gyration in reciprocal systems), but without the need of an externally imposed temperature gradient. 

Finite current without imposing any external gradient in a homogeneous, isotropic environment is a hallmark of living systems. Though fairly simple, Brownian dimer with suitable variations, have been used to model fundamental principles of widely varied phenomena in living systems. For example, Brownian inchworm \cite{kumar2008active,baule2008exact} to model the movement of helicases on DNA \cite{yu2006structure}, the crawling of keratocytes with treadmilling actin \cite{verkhovsky1999self}, chromatin remodeling with molecular motors \cite{blossey2019chromatin}, myosin-VI translocation and its load induced anchoring \cite{altman2004mechanism} and more generally, the walking of processive molecular motors on microtubules \cite{vale2000way}. Furthermore, a Brownian dimer has also been used to model enzyme and enzymatic reactions \cite{hosaka2020shear}, where non-reciprocal interactions can be present as well \cite{mandal2024molecular,sapre2025non}. Autonomous microscopic gyration without the aid of an external  temperature gradient but only due to the inherent non-reciprocal interaction within the monomers of a Brownian dimer can be an effective physical model to investigate the fundamental working principles of such phenomena in more generic, complex living systems.      
%%%%%%%%%%%%%%%%%%%%%%%%%%%%%%%%%%%%%%%%%%%%%%%%%%%%%%%

\section{Model : Confined non-reciprocal dimer on a plane}

We consider two point-like overdamped Brownian particles (``monomers'') with positions $\{{\bf r}_1(t),{\bf r}_2(t)\}\in\mathcal{R}^2$, each confined by its own isotropic harmonic trap (of stiffness $k_0$) centered at the origin, and coupled to each other by a harmonic spring of natural (rest) length $l$. The two monomers share a common heat bath at temperature $T$. Crucially, the spring is \emph{non-reciprocal} --- the restoring force it exerts on monomer 1 has a stiffness $k_1$, while the force it exerts on monomer 2 (by the same bond) has, in general, a \emph{different} stiffness $k_2(\neq k_1)$. Such a coupling cannot be derived from any scalar interaction potential $U({\bf r}_1,{\bf r}_2)$, and hence, explicitly violates the action-reaction principle \cite{ivlev2015statistical,loos2020irreversibility}.

The overdamped Langevin equations can now be formulated as:
\begin{align}
\gamma\dot{\bf r}_1 &= -k_0{\bf r}_1 - k_1\big(|{\bf r}_1-{\bf r}_2|-l\big)\,\hat{\bf r}_{12} + {\bm\xi}_1(t), \label{eq:le1}\\
\gamma\dot{\bf r}_2 &= -k_0{\bf r}_2 + k_2\big(|{\bf r}_1-{\bf r}_2|-l\big)\,\hat{\bf r}_{12} + {\bm\xi}_2(t), \label{eq:le2}
\end{align}
with, $\hat{\bf r}_{12}=(\bf r_1-\bf r_2)/|\bf r_1-\bf r_2|$, friction coefficient $\gamma$, and independent, zero-mean Gaussian white noises satisfying:
\begin{equation}
\langle \xi_{i,a}(t)\,\xi_{j,b}(t')\rangle = 2\gamma k_B T\,\delta_{ij}\delta_{ab}\,\delta(t-t'),
\end{equation}
with, $i,j\in\{1,2\}$ labeling the monomers and $a,b\in\{x,y\}$ the Cartesian components. Because both monomers are coupled to the \emph{same} bath, at the \emph{same} temperature, any non-equilibrium behavior found in our study can only be attributed to the non-reciprocity $(k_1\neq k_2)$ itself, and not to an external temperature gradient.

For $l\neq0$, the coupling force is a non-linear function of ${\bf r}_1,{\bf r}_2$ (as it depends on $|{\bf r}_1-{\bf r}_2|$), which mixes all the $x$- and $y$-components of the two monomers and precludes a closed-form stationary solution. For $l=0$, however, $(|{\bf r}_1-{\bf r}_2|-l)\hat{\bf r}_{12}$ reduces to a linear vector ${\bf r}_1-{\bf r}_2$, and Eqs.~(\ref{eq:le1})-(\ref{eq:le2}) become an exactly-solvable linear (multi-variate Ornstein-Uhlenbeck) system. We treat this case first, in full analytical detail, before turning to the finite-$l$ case numerically.

\section{Exact stationary distribution for zero rest length}
\label{sec:l0dist}

At $l=0$, the sets of co-ordinates $(x_1,x_2)$ and $(y_1,y_2)$ decouple, due to the isotropy of the trap and of the coupling, and they obey an identical closed $2\times2$ linear system. Denoting $\bm z=(x_1,x_2)^T$ (the same equations hold for $(y_1,y_2)$ as well), Eqs.~(\ref{eq:le1})-(\ref{eq:le2}) become:
\begin{equation}
\gamma\dot{\bm z} = -A\bm z + \bm\xi(t), \qquad A=\begin{pmatrix} k_0+k_1 & -k_1\\ -k_2 & k_0+k_2\end{pmatrix}.
\label{eq:linsys}
\end{equation}
Note that the drift matrix $A$ is manifestly \emph{not symmetric} unless $k_1=k_2$: this asymmetry is the direct signature of non-reciprocity, and is the sole reason the results shown below differ qualitatively from the equilibrium (reciprocal) dimer.

The stationary probability distribution of a linear Langevin system of the form~(\ref{eq:linsys}) is a zero-mean Gaussian distribution, $P(\bm z)\propto \exp\left(-\tfrac12 \bm z^T C^{-1}\bm z\right)$, with the steady-state covariance matrix $C=\langle \bm z\bm z^T\rangle$, obtained from the stationary Lyapunov equation \cite{gajic2008lyapunov}:
\begin{equation}
A C + C A^T = 2\gamma k_B T\,\mathbf{I}.
\label{eq:lyap}
\end{equation}
Equation~(\ref{eq:lyap}) is a linear algebraic equation for the three independent entries of the symmetric matrix, $C=\begin{pmatrix}p & q\\ q& r\end{pmatrix}$, and can be solved directly by matching the respective coefficients (see Supplementary Material for the algebraic details). The explicit moments are: 

%(brief method: substitute $C$ into Eq.~(\ref{eq:lyap}), which yields three linear equations for $p,q,r$; the full elimination is given in the Supplementary Material, Sec.~S1). 

\begin{align}
q &= \frac{k_B T\big[k_0(k_1+k_2)+k_1^2+k_2^2\big]}{k_0\,\kappa\,\Sigma}, \label{eq:q}\\[2pt]
p &= \frac{k_B T + k_1 q}{k_0+k_1}, \qquad r = \frac{k_B T + k_2 q}{k_0+k_2}, \label{eq:pr}
\end{align}
with, $\kappa \equiv k_0+k_1+k_2$, and, $\Sigma\equiv 2k_0+k_1+k_2$. The stationary joint probability density in the same-coordinate-different-monomer plane / shape plane $(x_1,x_2)$ (and, identically, in $(y_1,y_2)$) can then be written as:
\begin{equation}
P(x_1,x_2) = \frac{1}{2\pi\sqrt{\Delta}}\exp\!\left[-\frac{r x_1^2 - 2 q x_1 x_2 + p x_2^2}{2\Delta}\right],
\label{eq:Pxx}
\end{equation}
with, $\Delta \equiv pr-q^2$. Similarly, one can also obtain $P(y_1,y_2)$, and therefore, the overall composite distribution function becomes: $P(x_1,x_2,y_1,y_2)=P(x_1,x_2)P(y_1,y_2)$.  Needless to say that this factorization of the composite density into decoupled $x-$ and $y-$parts can be carried out only in the $l=0$ regime, where no dynamical coupling exists between the former and the latter whatsoever.

\section{Steady-state current and the role of non-reciprocity}
\label{sec:current}

\subsection{General analytical expressions}

For a linear system of the form~(\ref{eq:linsys}), having a stationary Gaussian density $P(\bm z)$, the Fokker-Planck probability current takes the compact form (details given in Supplementary Material, Sec.~S2):
\begin{equation}
\bm J(\bm z) = Q\,\bm z\, P(\bm z), \qquad Q = \frac{k_B T}{\gamma}C^{-1} - \frac{A}{\gamma}.
\label{eq:Jgeneral}
\end{equation}
The stationary state is in equilibrium (that is, zero current everywhere) if and only if $QC$ is symmetric; a non-zero anti-symmetric part of $QC$ is both necessary and sufficient for a genuine, non-vanishing steady-state current. Carrying out this calculation explicitly, we obtain the components of $\bm J$  as:
\begin{equation}
J_1(x_1,x_2) = \frac{w}{\gamma\Delta}\big(q x_1 - p x_2\big) P(x_1,x_2),
\label{eq:Jshape1}
\end{equation}
\begin{equation}
J_2(x_1,x_2) = \frac{w}{\gamma\Delta}\big(r x_1 - q x_2\big) P(x_1,x_2),
\label{eq:Jshape2}
\end{equation}
with the pre-factor,
\begin{equation}
w = \frac{k_B T\,(k_2-k_1)}{\Sigma}.
\label{eq:w}
\end{equation}
Equation~(\ref{eq:w}) is the pivotal result of our study: the entire current field is proportional to $(k_2-k_1)$. If $k_1=k_2$ (i.e. a reciprocal spring), we obtain $w\equiv0$ and $\bm J\equiv0$ everywhere --- the system is then genuinely in thermal equilibrium, as it must be, since a reciprocal linear coupling in a single-temperature bath is always derivable from a scalar potential. For \emph{any} $k_1\neq k_2$, however small the difference is, $w\neq0$ and a non-zero stationary current exists throughout the shape plane. This implies that non-reciprocity alone can drive the system arbitrarily far from equilibrium: there is no temperature gradient, no external force, and no explicit time-dependence anywhere in the model. Later, we will show that $\bm J$ is indeed a rotational current as we observe in microscopic Brownian gyration. The sign of $w$, and hence, the sense of circulation of $\bm J$, is fixed entirely by the sign of $k_2-k_1$. Similarly, one can compute $\bm J(y_1,y_2)$. 

From the equations of motion it is evident that though $(x_1,x_2)\in X $ (or, $(y_1,y_2)\in Y$) are coupled among themselves, $X$ and $Y$ are independent of each other. However, to obtain a finite current involving multiple degrees of freedom (DoF), we need to couple them. Here, $x_1$ and $x_2$ are coupled by the non-reciprocity of the inter-particle interaction, and that can also drive the system away from equilibrium, as shown above. Hence, ${\bm J (x_1,x_2)}$  becomes non-zero. Similarly, ${\bm J(y_1,y_2)}$ also becomes non-zero. However, there is no coupling between $X$ and $Y$, and therefore, the currents such as ${\bm J}(x_1,y_2)$, ${\bm J}(x_1,y_1)$, etc. are always zero. The persistent non-vanishing current found above exists in the plane obtained by pairing the same Cartesian component of two different monomers, $(x_1,x_2)$ or $(y_1,y_2)$, which is precisely the pair involved in non-reciprocal exchange of momentum between the two monomers, breaking action-reaction symmetry of the spring.

\vspace{-1 cm}
    
\begin{figure}[ht]
    \centering
    \includegraphics[width= \linewidth]{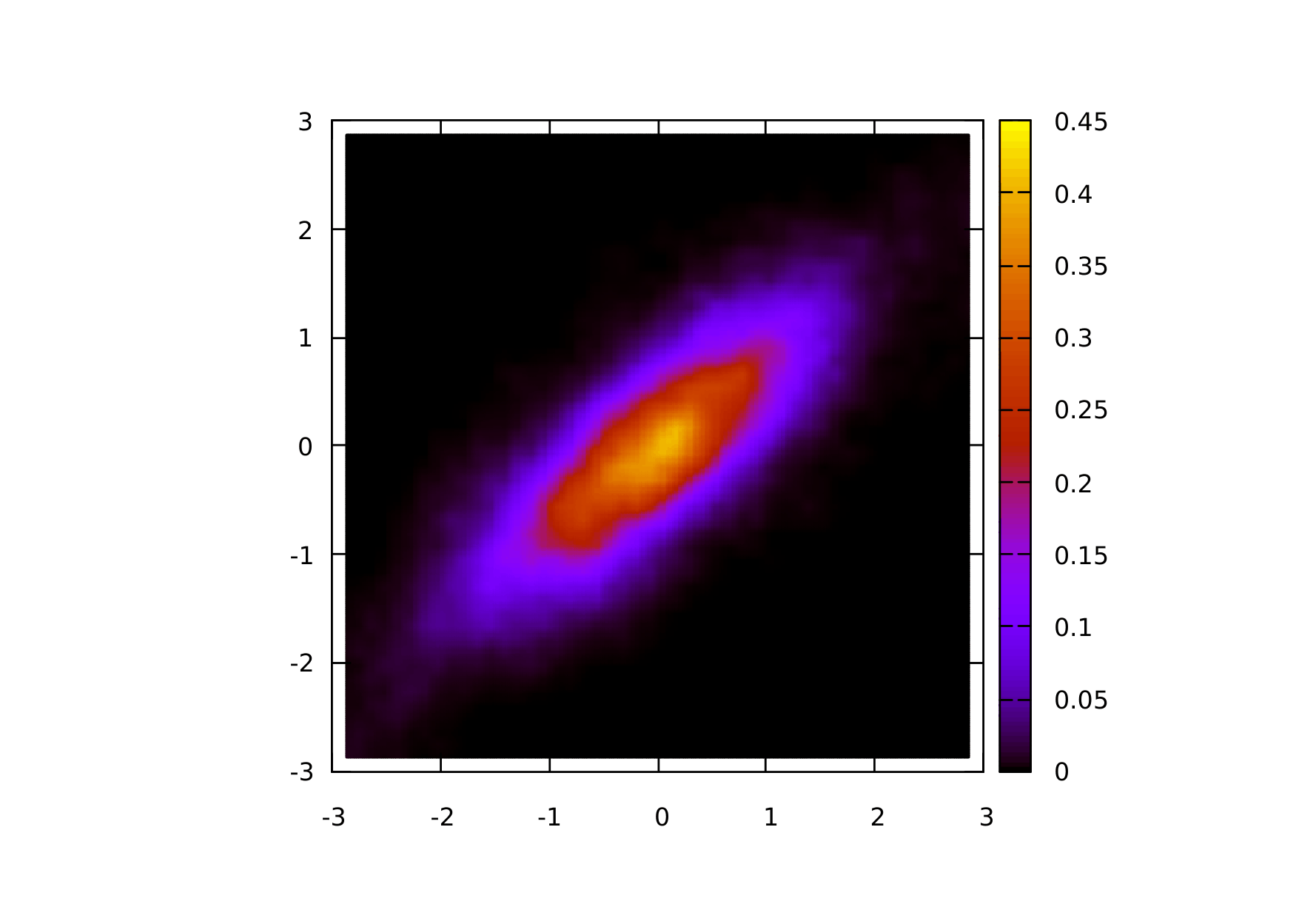}
    
      %\vspace{0.7 cm}
      
    \includegraphics[width= \linewidth]{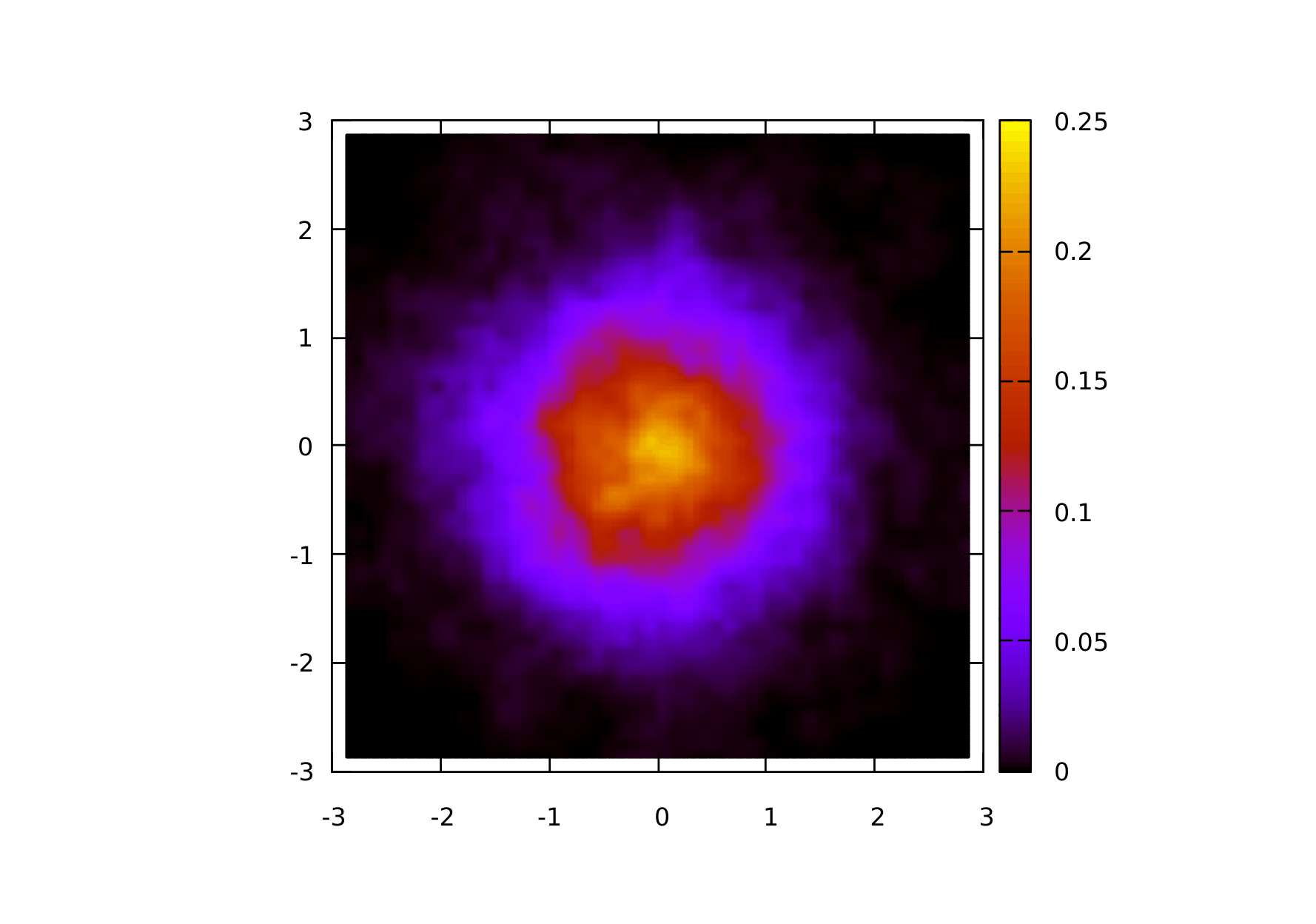}
   \caption{Heatmap of the joint probability distribution of two variates, obtained by numerically integrating the Langevin equations (for $l=0$): Top --- $P(x_1 , x_2)$ [identical for $\{y_1, y_2\}$], Bottom --- $P(x_1, y_1)$ [identical for $\{x_1, y_2\}$, etc.]. Note the distinctly different spread and alignment of the two distribution profiles. This is simply because:$P(x_1,y_1)=\int dx_2dy_2 P(x_1,x_2,y_1,y_2)=\int dx_2dy_2 P(x_1,x_2)P(y_1,y_2)=\int dx_2 P(x_1,x_2)\int dy_2 P(y_1,y_2)= P(x_1)P(y_1)$, and so on. However, this is not true for $P(x_1,x_2)$ or $P(y_1,y_2)$. The color scale indicates the value of probability. The values of the parameters used in simulation are: $k_0=1,T=1,\gamma=1,k_1 =1, k_2=5$.}
  \label{fig:probdist}
\end{figure}

\begin{figure}[ht]
    \centering
    \includegraphics[width= 1\linewidth]{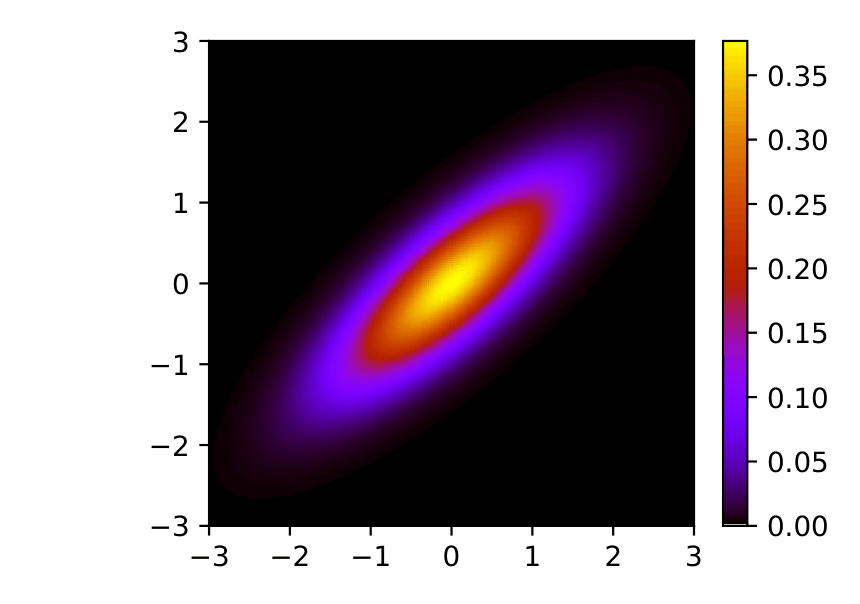}

    \vspace{1 cm}
    
    \includegraphics[width= 1\linewidth]{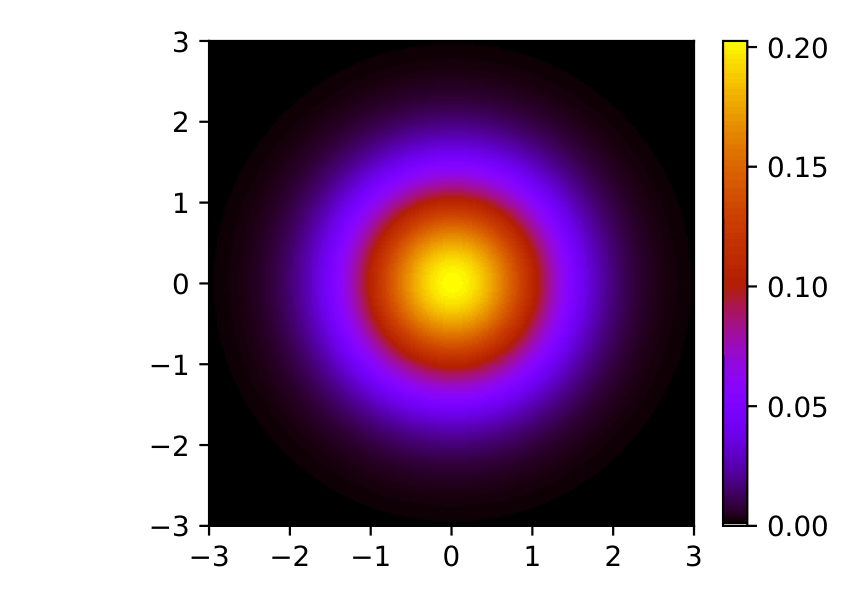}
   \caption{Analytical counterpart of Fig.~\ref{fig:probdist}, obtained from the closed-form Gaussian distribution $P(x_1,x_2)$ from Eq.~(\ref{eq:Pxx}), and its marginal distribution $P(x_1,y_1)=P(x_1)P(y_1)$ (for $l=0$): Top --- $P(x_1,x_2)$ [identical for $\{y_1,y_2\}$], Bottom --- $P(x_1,y_1)$ [identical for $\{x_1,y_2\}$, etc.]. Same color scale, axis range, and layout are kept as Fig.~\ref{fig:probdist}, to allow a direct visual comparison with the simulation. Parameter values are: $k_0=1,T=1,\gamma=1,k_1=1,k_2=5$.}
  \label{fig:probdist_analytic}
\end{figure}

\begin{figure}[H]
    \centering
    \includegraphics[width= \linewidth]{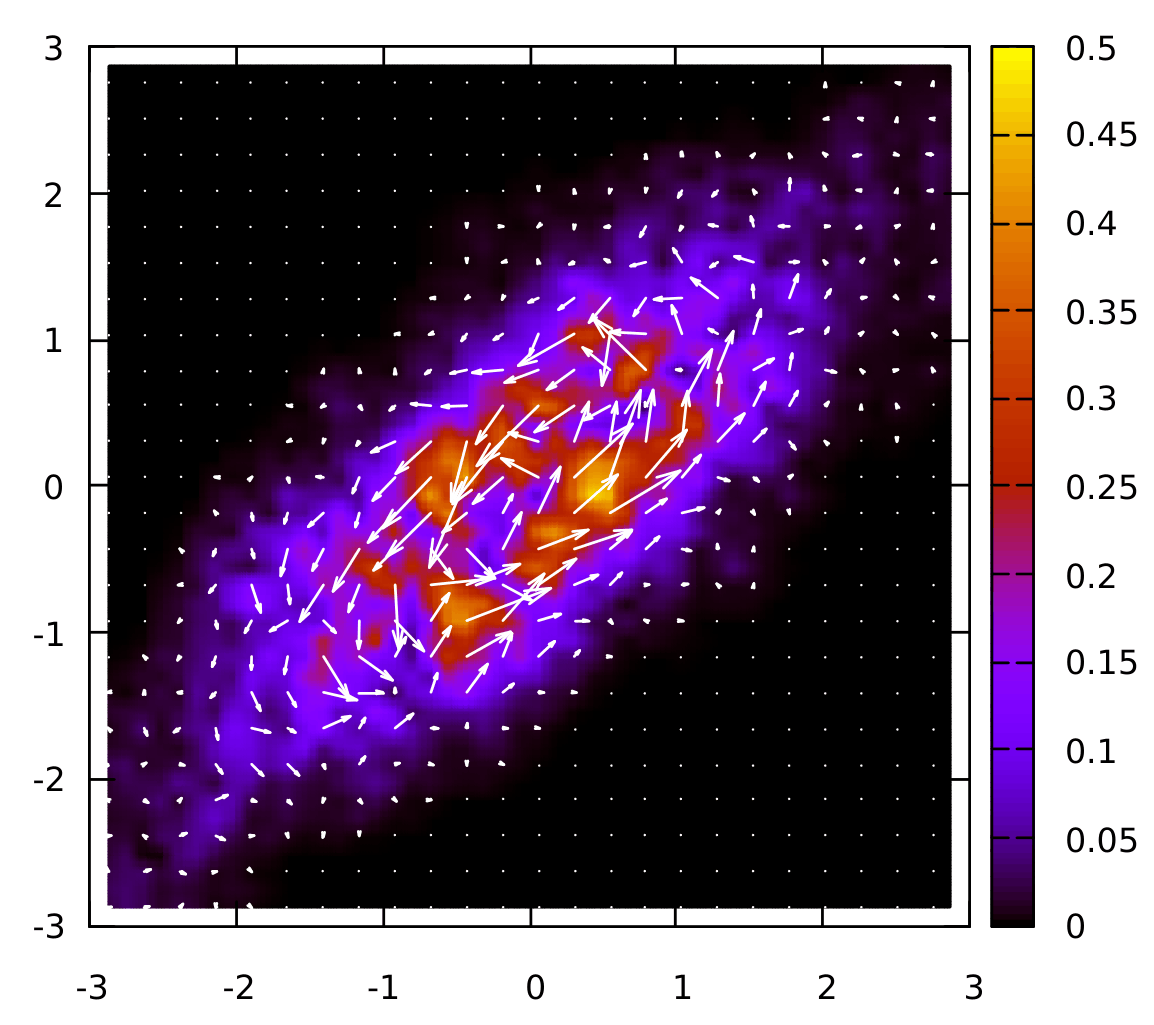}
    
    \vspace{-0.5 cm}
    
    \includegraphics[width= \linewidth]{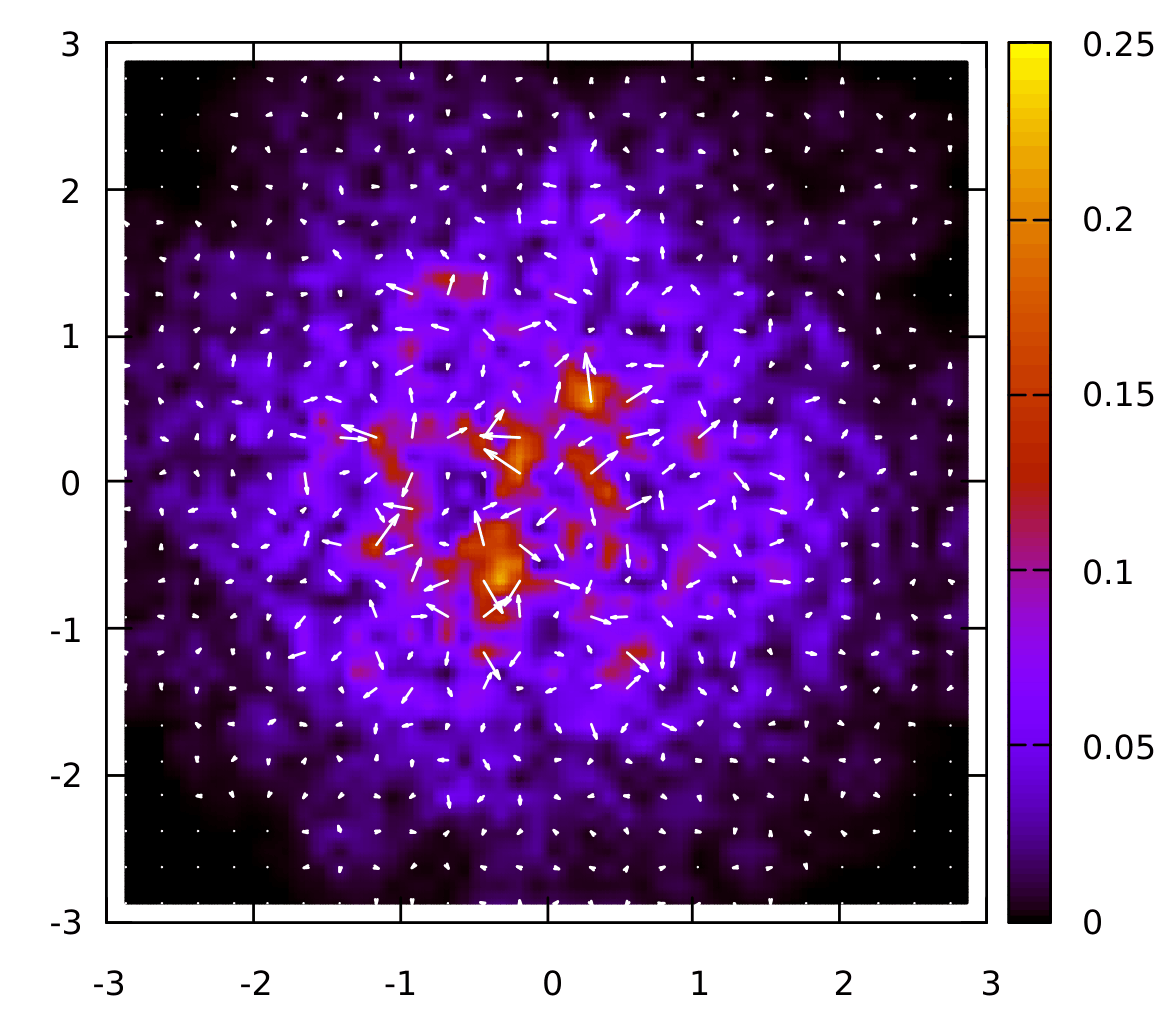}
    \caption{Heatmap of the magnitude of current of two variates with an overlay of the stream plot of the current vector, obtained by numerically integrating the Langevin equations (for $l=0$): Top --- in the $\{x_1, x_2\}$ plane [identical for $\{y_1, y_2\}$], Bottom --- in the $\{x_1, y_1\}$ plane [identical for $\{x_1, y_2\}$, etc.]. Note the absence of vorticity in the latter case. This implies the average current is non-zero in $x_1-x_2$ (or, $y_1-y_2$) plane but zero in $x_1-y_1$ (or, $x_2,y_2$) plane. The color scale indicates the magnitude of current, while the white arrows point in its direction. The parameter values used in simulation are kept the same as before.}
  \label{fig:current}
\end{figure}

\vspace{- 1 cm}

\begin{figure}[H]
    \centering
    \includegraphics[width= 0.9\linewidth]{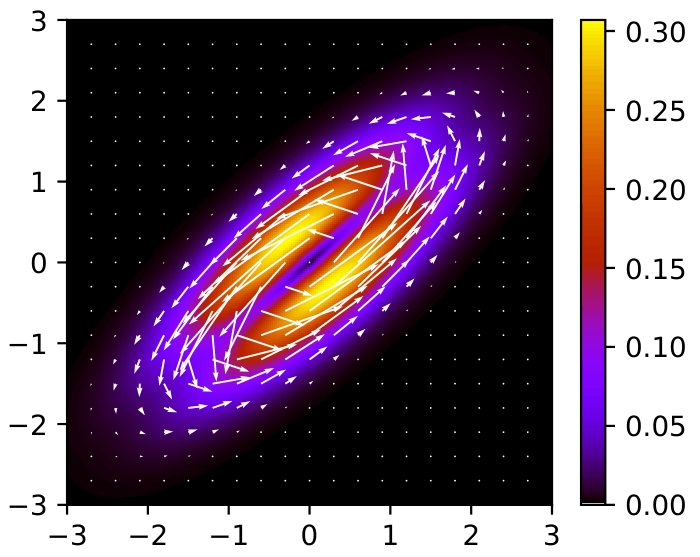}
    \caption{Analytical counterpart of the top panel of Fig.~\ref{fig:current}, obtained from the closed-form current in Eqs.~(\ref{eq:Jshape1})-(\ref{eq:Jshape2}) (for $l=0$) --- Plot of $\bm J(x_1,x_2)$ [identical for $\{y_1,y_2\}$], with white arrows showing the exact current direction. Same color scale, axis range, and layout are kept as Fig.~\ref{fig:current}. Parameter values are: $k_0=1,T=1,\gamma=1,k_1=1,k_2=5$.}
  \label{fig:current_analytic}
\end{figure}

\section{Curl of the current: Mapping with Brownian gyration}
\label{sec:curl}

\begin{figure}[H]
    \centering
    \includegraphics[width= \linewidth]{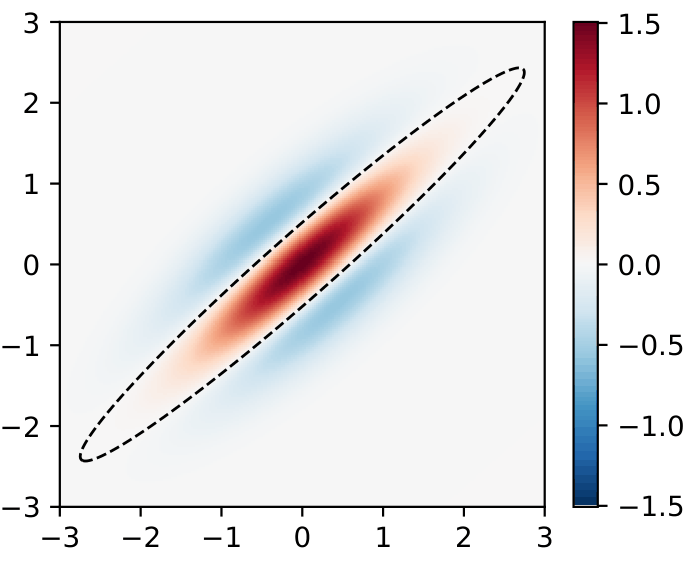}
    \caption{Analytical local vorticity $S(x_1,x_2)=(\nabla\times\bm J)_z$ (for $l=0$), obtained from Eq.~(\ref{eq:curl}), showing the central vortex core surrounded by a ring of opposite-sign vorticity (dashed line is the zero-vorticity contour, $Q(x_1,x_2)=\Delta(p+r)$). Parameter values are: $k_0=1,T=1,\gamma=1,k_1=1,k_2=5$.}
  \label{fig:curl_analytic}
\end{figure}

As the stationary current is divergence-free ($\nabla\cdot{\bm{J}}=0$), the local feature in two dimensions is its scalar curl, defined as:

\begin{equation}
S(x_1,x_2) \;\equiv\; \big(\nabla\times\bm J\big)_z \;=\; \frac{\partial J_2}{\partial x_1} - \frac{\partial J_1}{\partial x_2},
\label{eq:curldef}
\end{equation}
where, $S$ is called the \emph{local vorticity}, or the \emph{rotation-rate density}, of the stationary probability flow, and is the direct two-dimensional analogue of the ``gyration'' that characterizes the two-temperature Brownian gyrator (realized experimentally in \cite{argun2017experimental}). Carrying out the differentiation of Eqs.~(\ref{eq:Jshape1},\ref{eq:Jshape2}) explicitly (see Supplementary Material, Sec.~S3) yields a closed form of $S$:

\begin{equation}
S(x_1,x_2) = \frac{w}{\gamma\Delta}\,P(x_1,x_2)\left[(p+r) - \frac{Q(x_1,x_2)}{\Delta}\right],
\label{eq:curl}
\end{equation}

with, $Q(x_1,x_2) \equiv (r x_1 - q x_2)^2 + (q x_1-p x_2)^2$. At the center of the distribution, Eq.~(\ref{eq:curl}) gives:

\begin{equation}
S(0,0) = \frac{w\,(p+r)}{\gamma\Delta}\,P(0,0),
\label{eq:S0}
\end{equation}

which indicates a natural local angular frequency of rotation of the probability flow, $\omega_0\equiv S(0,0)/P(0,0)=w(p+r)/(\gamma\Delta)$. As $Q(x_1,x_2)$ is a positive-definite quadratic form while $(p+r)$ is a constant, Eq.~(\ref{eq:curl}) changes sign on the ellipse defined by: $Q(x_1,x_2)=\Delta(p+r)$ --- the vorticity field, therefore, consists of a central core of one sign of rotation surrounded by a ring of opposite sign, i.e.\ a bound vortex-antivortex pair, before a fast Gaussian decaying to zero far from the origin. This structure is shown in Fig.~\ref{fig:curl_analytic}.

\begin{figure}[H]
    \centering
    \includegraphics[width=1.1\linewidth]{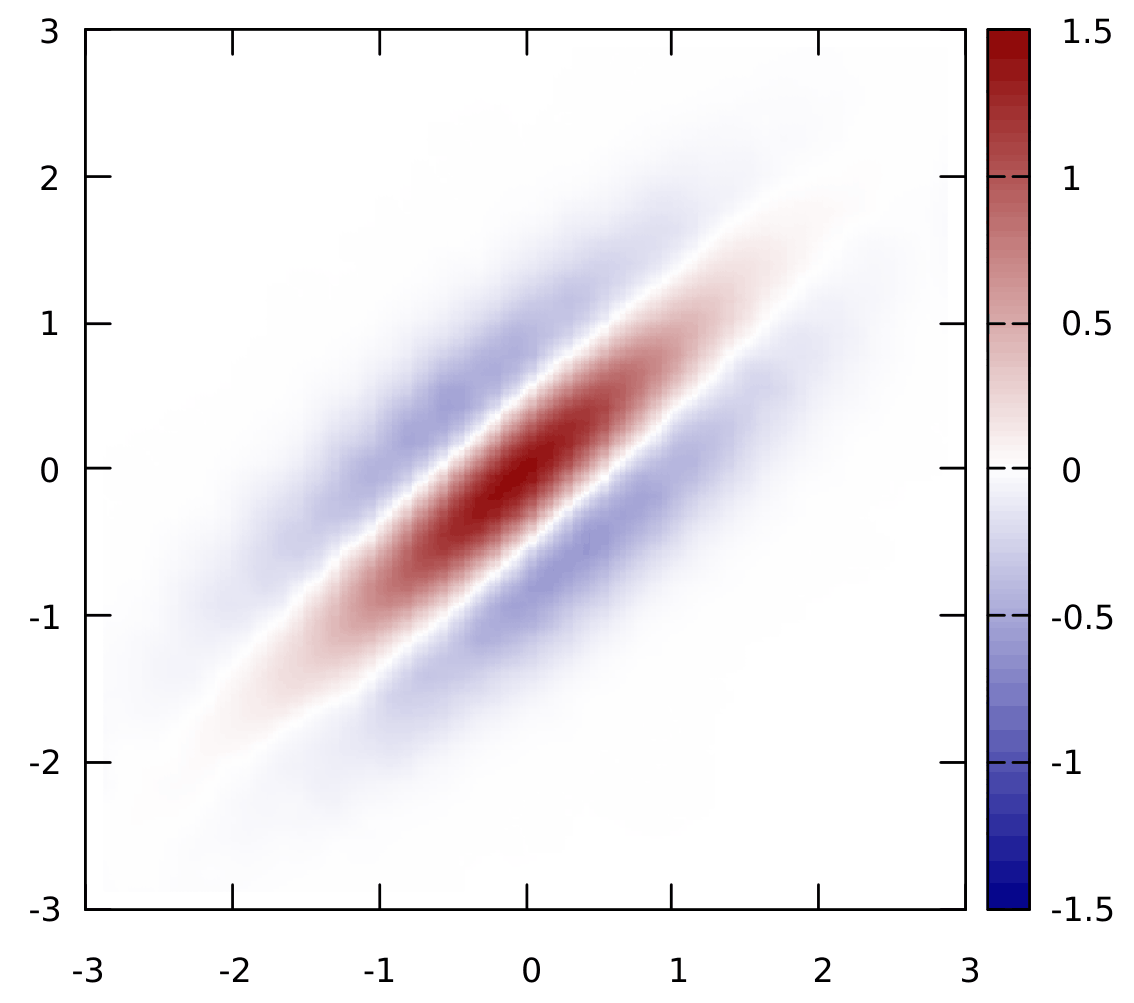}
    \caption{Heatmap of the vorticity $S(x_1 , x_2)$, obtained by numerically integrating the Langevin equations (for $l=0$). Note the distinctly separated vortex-antivortex regions, demarcated by a zero-vorticity boundary. The color scale indicates the value of vorticity. The parameter values used in simulation are kept the same as before.}
  \label{fig:curl}
\end{figure}

%\vspace{-2 cm}

Equation~(\ref{eq:curl}) precisely makes a connection to microscopic gyration and the mapping between these two models becomes more evident. Both the Brownian gyrator and the present non-reciprocal dimer are linear Ornstein-Uhlenbeck systems with a Gaussian stationary state and a current of the generic form $\bm J = Q\bm z P(\bm z)$; a non-zero curl of $\bm J$, and hence gyration, requires only that $QC$ must possess a non-zero anti-symmetric part. What distinguishes the two systems is \emph{how} that anti-symmetry is generated. In the Brownian gyrator, the drift (or, coupling) matrix $A$ is symmetric (as the spring is reciprocal) and the detailed balance is broken by making the diffusion matrix anisotropic, i.e.\ by coupling the two degrees of freedom to \emph{different} temperatures $(T_1\neq T_2)$ \cite{mancois2018two}. In the present model, however, the diffusion matrix is isotropic (both monomers share a single bath at a uniform temperature $T$) but the system is driven out of equilibrium by making $A$ itself asymmetric, as $k_1\neq k_2$. These are the two independent ways to achieve microscopic gyration. Non-reciprocity, therefore, provides a mechanically-realizable, single-temperature route to exactly the same class of gyrating non-equilibrium steady-states which are usually associated with explicit temperature gradients.

%of violating the generalised detailed-balance condition $AD=DA^T$. 

%for a linear system with diffusion matrix $D$, and Loos and Klapp~\cite{LoosKlapp2020} showed that, for a pair of linearly coupled degrees of freedom, the two mechanisms can in fact be mapped onto one another by a linear rescaling of variables. 

\section{Numerical results for finite rest length}
\label{sec:finitel}

\begin{figure}[H]
    \centering
    \includegraphics[width= 1.1\linewidth]{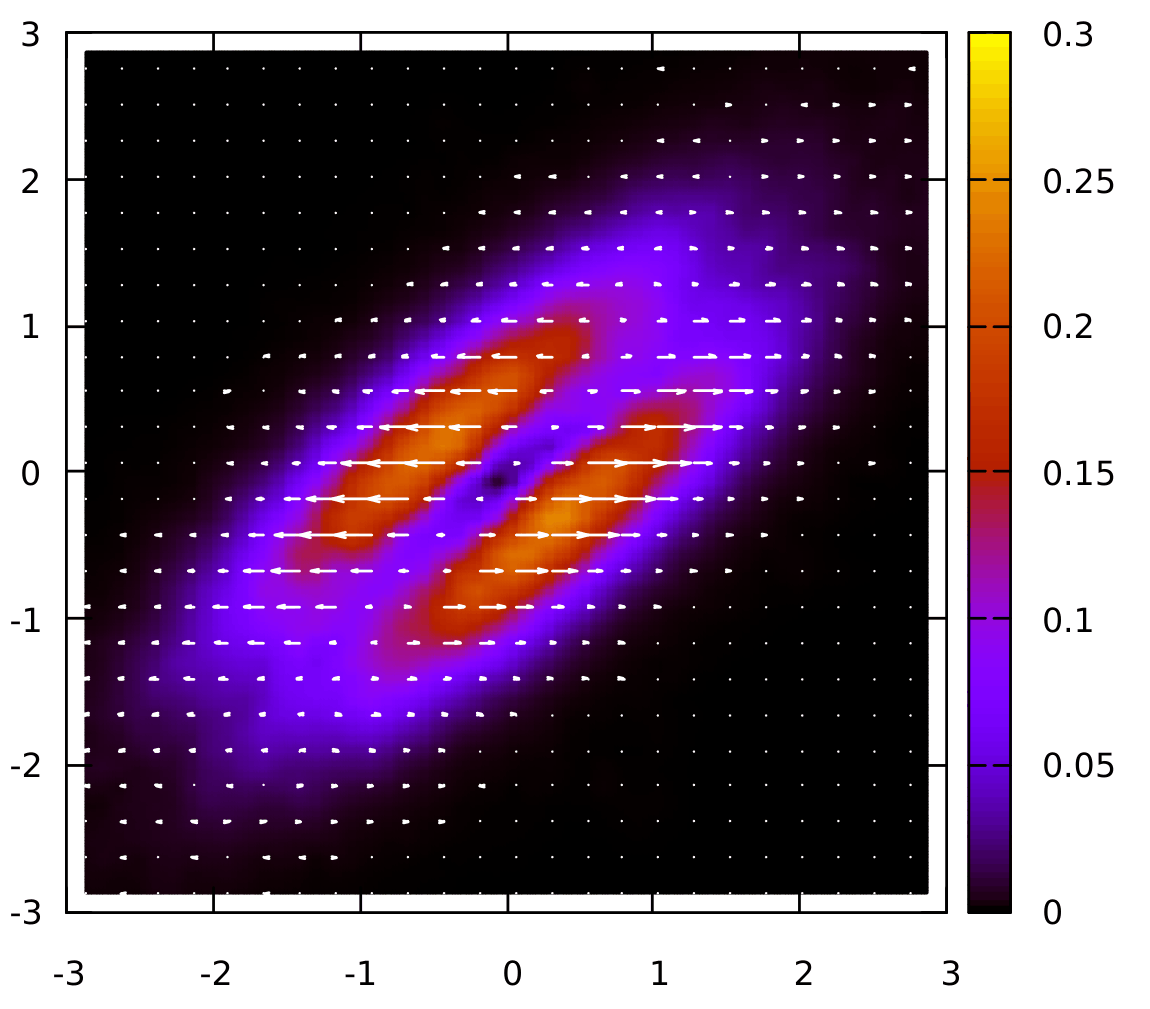}

    \vspace{0.5 cm}
    
    \includegraphics[width= 1.1\linewidth]{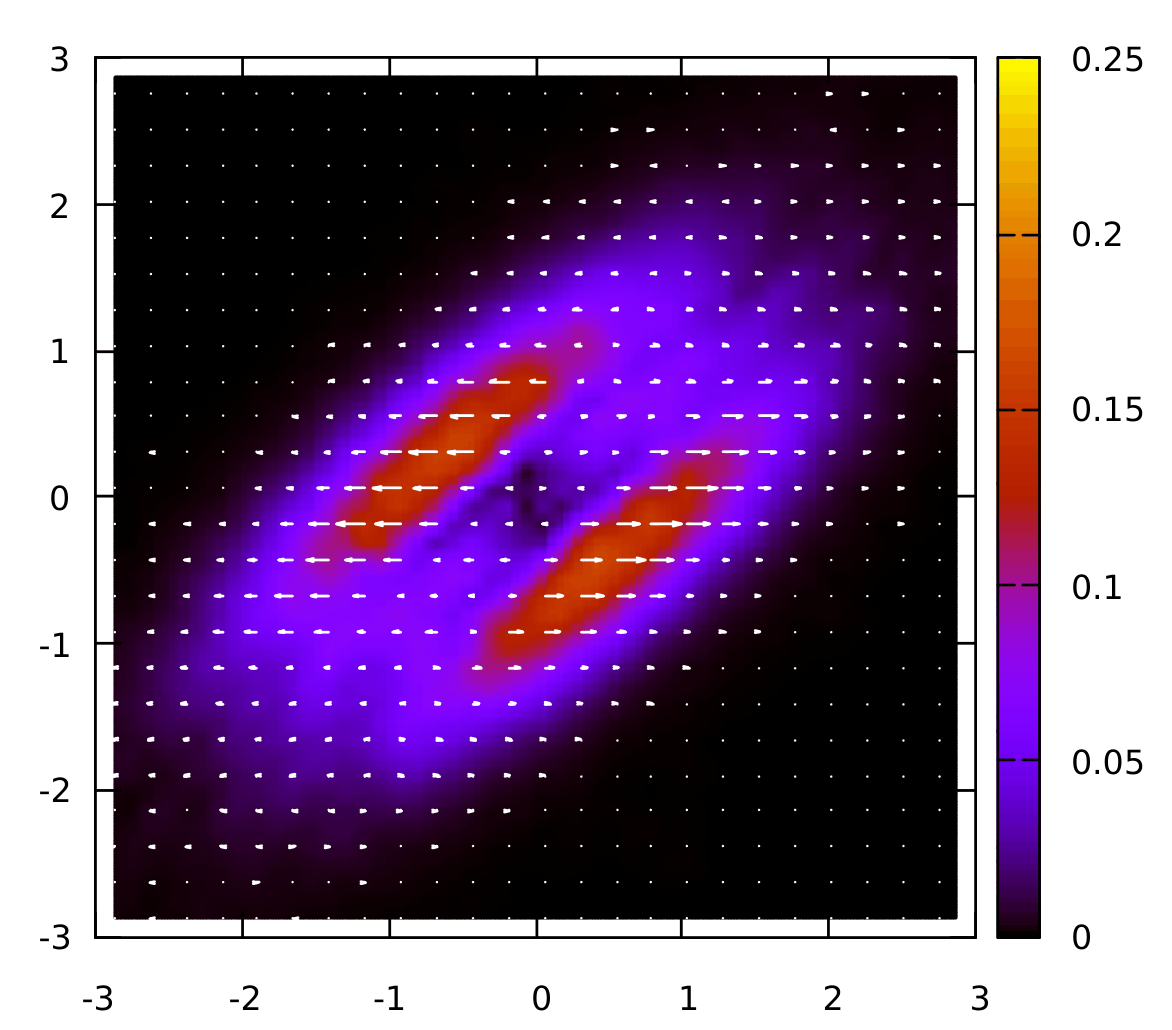}
    \caption{Heatmap of the magnitude of current of two variates $\{x_1, x_2\}$ with an overlay of the stream plot of the current vector, obtained by numerically integrating the Langevin equations (for $l\neq0$): Top --- $l=0.2$, Bottom --- $l=0.5$. Note the extent of separation of peak currents in the two cases. The color scale indicates the magnitude of current, while the white arrows point in its direction. The parameter values used in simulation are kept the same as before.}
  \label{fig:current_non-zerol}
\end{figure}

\begin{figure}[ht]
    \centering
    \includegraphics[width=1.1\linewidth]{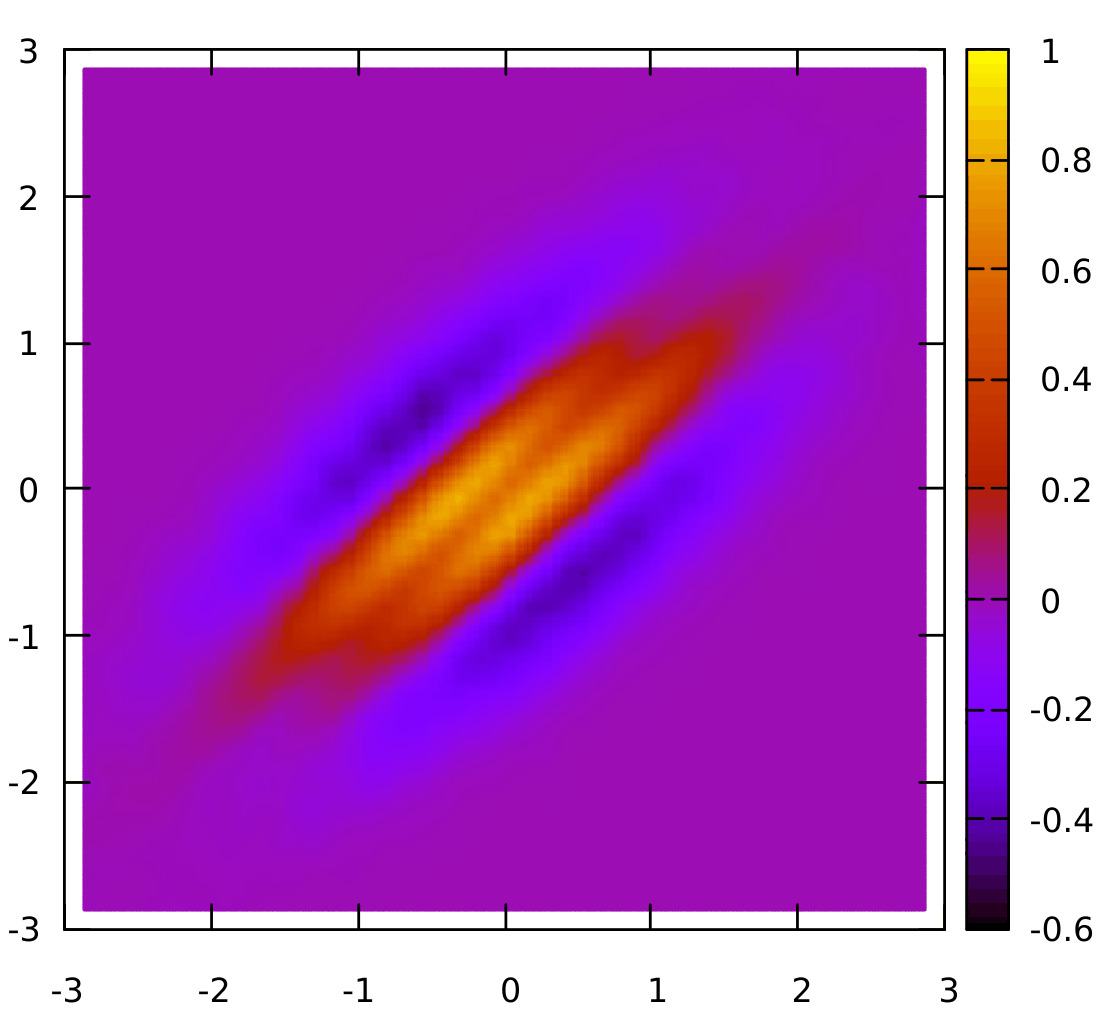}

    \vspace{0.5 cm}
    
    \includegraphics[width=1.1\linewidth]{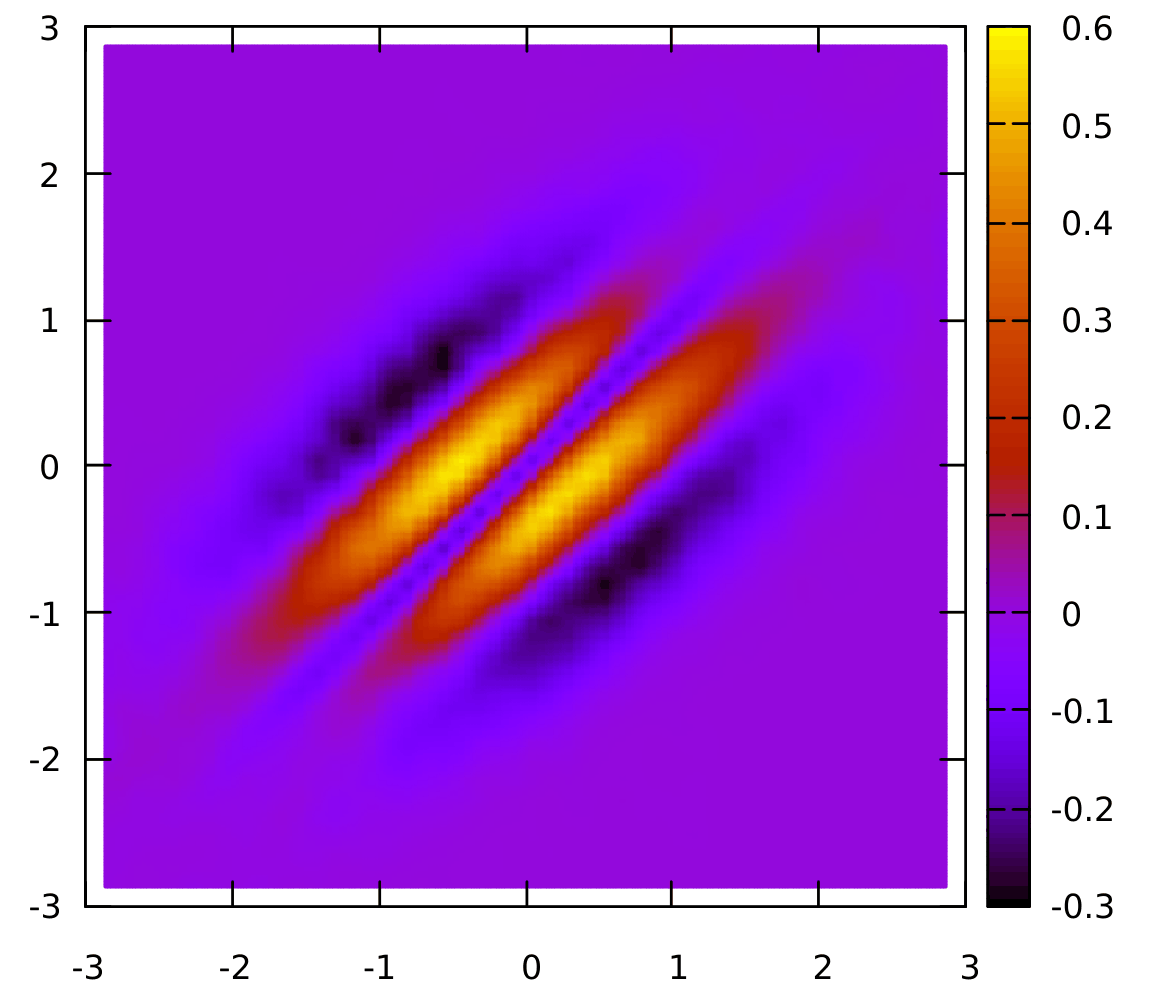}
    \caption{Heatmap of the vorticity $S(x_1 , x_2)$, obtained by numerically integrating the Langevin equations (for $l\neq0$): Top --- $l=0.2$, Bottom --- $l=0.5$. Note the different arrangements of the vortex-antivortex regions in the two cases. The color scale indicates the value of vorticity. The parameter values used in simulation are kept the same as before.}
  \label{fig:curl_non-zerol}
\end{figure}

For $l\neq 0$, the coupling force $k_i(|{\bf r}_1-{\bf r_2}|-l)\hat{\bf r}_{12}$ depends on $|{\bf r_1}-{\bf r_2}|=\sqrt{(x_1-x_2)^2+(y_1-y_2)^2}$, which couples the $X$- and $Y$-sets of co-ordinates and makes Eqs.~(\ref{eq:le1})-(\ref{eq:le2}) non-linear. There is no closed-form, stationary probability distribution for such model equations. We therefore integrate the full two-dimensional, four-variable, non-linear Langevin dynamics numerically to obtain the overall composite probability distribution, corresponding currents (Fig.[\ref{fig:current_non-zerol}]) and its vorticity (Fig.[\ref{fig:curl_non-zerol}]). Due to the non-zero rest-length $(l)$ of the spring, the current and vorticity fields are modified qualitatively. The current density (magnitude of the current vector) profile exhibits two distinct lobes, the distance between which increases as $l$ increases. For $l=0$ case, the current was primarily rotational, whereas for the $l>0$ case, the current has a translational component as well. In the vorticity field, we observe that the number of vortex-antivortex pairs has been increased as $l$ becomes non-zero.

\vspace{-0.3 cm}

\section{Coupling to two different temperatures: Interplay of mechanical and thermal asymmetries}

All the above results can be generalized by coupling the individual monomers to \emph{different} heat baths at temperatures $T_1\neq T_2$, i.e.\ to replace the noise correlator of Eqs.~(\ref{eq:le1})-(\ref{eq:le2}) by:
\begin{equation}
\langle \xi_{i,a}(t)\,\xi_{j,b}(t')\rangle = 2\gamma k_B T_i\,\delta_{ij}\delta_{ab}\,\delta(t-t'),
\label{eq:twoTnoise}
\end{equation}
while retaining $k_1\neq k_2$. For $l=0$, the dynamics is still linear and the stationary Lyapunov equation now reads:
\begin{equation}
AC+CA^T = 2D, \qquad D = \begin{pmatrix}k_B T_1 & 0\\ 0 & k_B T_2\end{pmatrix},
\label{eq:twoTlyap}
\end{equation}
which generalizes Eq.~(\ref{eq:lyap}) (with $D=k_B T\,\mathbf{I}$ being the special, equal-temperature case). Solving Eq.~(\ref{eq:twoTlyap}) exactly as in Sec.~\ref{sec:l0dist} (see Supplementary Material for details) gives:
\begin{equation}
q = \frac{k_2(k_0+k_2)\,k_B T_1 + k_1(k_0+k_1)\,k_B T_2}{k_0\,\kappa\,\Sigma},
\label{eq:twoTq}
\end{equation}
\begin{equation}
p = \frac{k_B T_1 + k_1 q}{k_0+k_1}, \qquad r = \frac{k_B T_2 + k_2 q}{k_0+k_2},
\label{eq:twoTpr}
\end{equation}
with $\kappa,\Sigma$ as before, so that the stationary distribution retains exactly the Gaussian form of Eq.~(\ref{eq:Pxx}), $P(x_1,x_2)=(2\pi\sqrt{\Delta})^{-1}\exp[-(rx_1^2-2qx_1x_2+px_2^2)/2\Delta]$, $\Delta=pr-q^2$, now with $p,q,r$ given by Eqs.~(\ref{eq:twoTq})-(\ref{eq:twoTpr}). 

%Remarkably, the algebra proceeds exactly as in Sec.~\ref{sec:current}: the antisymmetry argument used there to show $QC+(QC)^T=0$ did not actually require $D\propto\mathbf{I}$, only that $D$ be symmetric, so the current retains the identical functional form of Eqs.~(\ref{eq:Jshape}),

The analytics proceed similarly as before and we obtain: 
\begin{align}
J_1(x_1,x_2) &= \frac{w}{\gamma\Delta}(qx_1-px_2)\,P(x_1,x_2), \notag\\
J_2(x_1,x_2) &= \frac{w}{\gamma\Delta}(rx_1-qx_2)\,P(x_1,x_2), \label{eq:twoTJ}
\end{align}
with the generalized pre-factor,
\begin{equation}
w = \frac{k_B\big(k_2 T_1 - k_1 T_2\big)}{\Sigma}.
\label{eq:twoTw}
\end{equation}
Equation~(\ref{eq:twoTw}) unifies the two mechanisms for generating a Brownian-gyrator-type current in a linear system: setting $T_1=T_2=T$ recovers Eq.~(\ref{eq:w}), where current is driven purely by non-reciprocity ($k_1\neq k_2$); setting instead $k_1=k_2=k$ (i.e. a reciprocal spring) gives $w=k_B k(T_1-T_2)/\Sigma$, the current now being driven purely by the temperature difference --- the mechanism of the original Brownian gyrator. In general, the current (and hence, the gyration), vanishes if $k_2T_1=k_1T_2$. Non-reciprocity and temperature gradient are therefore, in the context of steady-state gyration, interchangeable. They can even be suitably tuned to cancel one another exactly for a matched $k_1/k_2$ and $T_1/T_2$.

\section{Concluding remarks and further prospects} To summarize, we have shown that a two-dimensional dimer consisting of overdamped Brownian monomers held in a common isotropic trap and in contact with a single heat bath, can go arbitrarily far from equilibrium and produce non-zero, steady-state currents, when the monomers interact with each other by a non-reciprocal harmonic spring of length $l$. For $l=0$, the system is linear and we have analytically obtained the exact stationary joint distribution $P(x_1,x_2)$, the associated closed-form steady-state current and its vorticity, and drew parallels with the phenomenon of Brownian gyration. We have also generalized these results when the two monomers are kept at different temperatures. For $l\neq 0$, all these quantities are calculated numerically. These results illustrate that a local violation of Newton's third law (in the effective interactions between constituent elements) can actually drive the system far from equilibrium, even without any temperature difference between them.

This can be realized in synthetic active systems such as feedback-controlled active colloids. These results may also help us to understand the source of activity in living systems, which are inherently non-reciprocal. One can further develop the model considering the spring to be anisotropic, which will now couple $\{x_i\}$ and $\{y_i\}$, and therefore the current that depends on these coordinates will be non-zero. It would also be interesting to explore the molecular origin of non-reciprocity in enzymatic reactions \cite{mandal2024molecular,sapre2025non} with such simple non-reciprocal models.

\begin{acknowledgments}
S.P. acknowledges the hospitality of Stat. Mech. Meet Kolkata (SMMK-2024) organized by the S.N. Bose National Centre for Basic Sciences, where the current work was presented and discussed.    S.D. acknowledges University Grants Commission (UGC), India and University of Calcutta for financial assistance. A.S. acknowledges support from Anusandhan National Research Foundation (ANRF), DST (India), under the Advanced Research Grant (ARG) scheme (No.- DST(IN), ANRF/ARG/2025/009343/PS). A.S. also acknowledges support from CY Initiative of Excellence (Grant: “Investissements dAvenir” ANR-16-IDEX-0008) under CY Advanced Studies (CYAS), where the present work was partially developed. 
\end{acknowledgments}

 \bibliographystyle{unsrt}
\bibliography{reference} 

\vspace{10 em} 

\begin{widetext}

\section*{Supplementary material --- Details of some non-trivial calculations}

This end matter provides the derivations summarized in the main paper. We have considered $\bm z=(x_1,x_2)^T$ at $l=0$, where the dynamics obey the linear Langevin equation:

\begin{equation}
\gamma\dot{\bm z} = -A\bm z + \bm\xi(t), \qquad A=\begin{pmatrix} k_0+k_1 & -k_1\\ -k_2 & k_0+k_2\end{pmatrix},
\tag{S0}
\end{equation}
with $\langle\xi_a(t)\xi_b(t')\rangle = 2\gamma k_B T\,\delta_{ab}\,\delta(t-t')$. Without any loss of generality, we set $\gamma=1$ (and can be restored by setting $k_B T\to k_B T/\gamma$, with an overall $1/\gamma$ in the current).

\section*{S1. Stationary covariances from the Lyapunov equation}

The stationary distribution of Eq.~(S0) is a zero-mean Gaussian distribution, $P(\bm z)\propto\exp(-\tfrac12\bm z^T C^{-1}\bm z)$, whose covariance $C=\begin{pmatrix}p&q\\q&r\end{pmatrix}$ obeys:

\begin{equation}
AC+CA^T = 2k_B T\,\mathbf{I}.
\tag{S1}
\end{equation}

The three independent components of Eq.~(S1) with $A$ from Eq.~(S0) can now be explicitly written as,

\begin{align}
(1,1):&\quad (k_0+k_1)\,p - k_1\,q = k_B T, \tag{S2a}\\
(2,2):&\quad -k_2\,q + (k_0+k_2)\,r = k_B T, \tag{S2b}\\
(1,2)[\text{same as } (2,1)]:&\quad (2k_0+k_1+k_2)\,q = k_1 r + k_2 p. \tag{S2c}
\end{align}

Equations (S2a) and (S2b) give $p$ and $r$ directly in terms of $q$,
\begin{equation}
p = \frac{k_B T+k_1 q}{k_0+k_1}, \qquad r = \frac{k_B T + k_2 q}{k_0+k_2}. \tag{S3}
\end{equation}

Substituting Eq.~(S3) into Eq.~(S2c) and multiplying throughout by $(k_0+k_1)(k_0+k_2)$ gives,

\begin{equation}
\Sigma\, q\,(k_0+k_1)(k_0+k_2) = k_B T\big[k_1(k_0+k_1)+k_2(k_0+k_2)\big] + q\,k_1k_2\,\Sigma,
\tag{S4}
\end{equation}

with, $\Sigma\equiv 2k_0+k_1+k_2$. Using the identity $(k_0+k_1)(k_0+k_2)-k_1k_2 = k_0(k_0+k_1+k_2)\equiv k_0\kappa$ (with, $\kappa\equiv k_0+k_1+k_2$), Eq.~(S4) simplifies to,

\begin{equation}
\Sigma\,\kappa\,k_0\, q = k_B T\big[k_0(k_1+k_2)+k_1^2+k_2^2\big],
\tag{S5}
\end{equation}

i.e.\ Eq.~(6) of the main text, $q = k_B T[k_0(k_1+k_2)+k_1^2+k_2^2]/(k_0\kappa\Sigma)$. Substituting back into Eq.~(S3) then gives $p$ and $r$ as quoted in the main text.

\section*{S2. Stationary current for the linear system}

For the SDE, $\dot{\bm z}=-A\bm z+\bm\eta$, with $\langle\eta_a(t)\eta_b(t')\rangle=2k_B T\,\delta_{ab}\delta(t-t')$, the Fokker-Planck equation becomes: $\partial_t P = -\nabla\cdot\bm J$, where:

\begin{equation}
\bm J(\bm z) = -A\bm z\,P(\bm z) - k_B T\,\nabla P(\bm z).
\tag{S6}
\end{equation}

For the stationary Gaussian distribution, $P(\bm z)\propto e^{-\bm (z^TC^{-1}\bm z)/2}$, one has $\nabla P = -C^{-1}\bm z\,P$, so Eq.~(S6) becomes:

\begin{equation}
\bm J(\bm z) = Q\bm z\,P(\bm z), \qquad Q \equiv k_B T\,C^{-1}-A.
\tag{S7}
\end{equation}

From the Lyapunov equation (S1), we know that $QC$ is \emph{always} antisymmetric, since $CA^T = 2k_B T\mathbf{I}-AC$ [as seen from Eq.~(S1)]:

\begin{align}
QC &= k_B T\,\mathbf{I} - AC, \notag\\
(QC)^T &= k_B T\,\mathbf{I} - CA^T = k_B T\,\mathbf{I}-\big(2k_B T\,\mathbf{I}-AC\big) = -k_B T\,\mathbf{I}+AC, \notag
\end{align}

so that $QC+(QC)^T=0$ identically, for \emph{any} $A$ (reciprocal or not), satisfying its own Lyapunov equation. The current is therefore zero everywhere if the (antisymmetric) matrix $QC$ vanishes, i.e.\ $QC=0$.

Writing the general $2\times2$ antisymmetric matrix as $QC = \tilde w\begin{pmatrix}0&-1\\1&0\end{pmatrix}$, evaluating $(QC)_{21} = -(AC)_{21} = k_2 p-(k_0+k_2)q$ using $A$ from Eq.~(S0) and $p,q,r$ from Sec.~S1, and simplifying with Eqs.~(S3) and (S5), one finds:

\begin{equation}
\tilde w = \frac{k_B T\,(k_2-k_1)}{\Sigma} \equiv w,
\tag{S8}
\end{equation}

which is zero when $k_1=k_2$. Inverting $Q=(QC)C^{-1}=w\begin{pmatrix}0&-1\\1&0\end{pmatrix}C^{-1}$ and using $C^{-1}=\Delta^{-1}\begin{pmatrix}r&-q\\-q&p\end{pmatrix}$ gives:

\begin{equation}
Q = \frac{w}{\Delta}\begin{pmatrix} q & -p\\ r & -q\end{pmatrix},
\tag{S9}
\end{equation}

and hence, from Eq.~(S7), the currents are (as quoted in Eqs.~(10)-(11) of the main text):

\begin{align}
J_1(x_1,x_2) &= \frac{w}{\Delta}\big(qx_1-px_2\big)P(x_1,x_2), \tag{S10a}\\
J_2(x_1,x_2) &= \frac{w}{\Delta}\big(rx_1-qx_2\big)P(x_1,x_2). \tag{S10b}
\end{align}

We verified Eqs.~(S8)-(S10) both by evaluating $A,C,Q$ for the reference parameter set of the main text and from numerical simulations (Fig.~1,3 of the main text).

\section*{S3. Curl of the current}

The magnitude of curl, $S=\partial_1 J_2-\partial_2 J_1$, requires the derivatives of the Gaussian distribution, $P(x_1,x_2)=\frac{1}{2\pi\sqrt\Delta}\exp\!\big[-(rx_1^2-2qx_1x_2+px_2^2)/(2\Delta)\big]$:

\begin{equation}
\partial_1 P = -\frac{r x_1-qx_2}{\Delta}\,P, \qquad \partial_2 P = -\frac{p x_2-qx_1}{\Delta}\,P.
\tag{S11}
\end{equation}

Differentiating Eq.~(S10b) with respect to $x_1$,

\begin{align}
\partial_1 J_2 &= \frac{w}{\Delta}\Big[r\,P + (rx_1-qx_2)\,\partial_1P\Big] \notag\\
&= \frac{w}{\Delta}\,P\left[r - \frac{(rx_1-qx_2)^2}{\Delta}\right],
\tag{S12}
\end{align}

and differentiating Eq.~(S10a) with respect to $x_2$,

\begin{align}
\partial_2 J_1 &= \frac{w}{\Delta}\Big[-p\,P + (qx_1-px_2)\,\partial_2P\Big] \notag\\
&= \frac{w}{\Delta}\,P\left[-p + \frac{(qx_1-px_2)^2}{\Delta}\right].
\tag{S13}
\end{align}

Subtracting Eq.~(S13) from Eq.~(S12) gives the closed form quoted as Eq.~(14) of the main text:

\begin{equation}
S(x_1,x_2) = \frac{w}{\Delta}\,P(x_1,x_2)\left[(p+r) - \frac{Q(x_1,x_2)}{\Delta}\right],
\tag{S14}
\end{equation}

with $Q(x_1,x_2)\equiv(rx_1-qx_2)^2+(qx_1-px_2)^2 = \Delta^2\,\bm z^T\!\left(C^{-1}\right)^{\!2}\bm z$ (a positive-definite quadratic form, since $C^{-1}$ is positive definite). We verified Eq.~(S14) by the numerical  evaluation of $\partial_1J_2-\partial_2J_1$ using Eqs.~(S10a)-(S10b) at several representative points $(x_1,x_2)$, finding an agreement with the closed form.

As $Q(x_1,x_2)$ is a positive-definite quadratic form while $\Delta(p+r)$ is a fixed constant, Eq.~(S14) changes sign exactly once along any ray from the origin, on the ellipse defined by: $Q(x_1,x_2)=\Delta(p+r)$. At the origin, $S(0,0) = w(p+r)P(0,0)/\Delta$, which sets the overall scale and sign of the central vortex core seen in Fig.~5 of the main text; the ring of opposite-sign vorticity surrounding it, and the subsequent decay of the magnitude of vorticity at large $|\bm z|$, both follow directly from Eq.~(S14).

\section*{S4. Generalization to two unequal temperatures}

Here we give the derivation of Eqs.~(16)-(21) of the main text, obtained by letting the two monomers be coupled to separate baths at temperatures $T_1\neq T_2$, i.e.\ $\langle\xi_{i,a}(t)\xi_{j,b}(t')\rangle=2\gamma k_B T_i\,\delta_{ij}\delta_{ab}\,\delta(t-t')$. We can redo the above algebra by keeping the diffusion matrix generic and diagonal, $D=\mathrm{diag}(k_B T_1,k_B T_2)$. This is a direct consequence of introducing a thermal anisotropy in the noise covariance.

\subsection*{S4.1 Stationary covariances}

The stationary Lyapunov equation is now $AC+CA^T=2D$, with $A$ as in Eq.~(S0). Writing out its three independent components with $D_1\equiv k_B T_1$, $D_2\equiv k_B T_2$:

\begin{align}
(1,1):&\quad (k_0+k_1)\,p - k_1\,q = D_1, \tag{S15a}\\
(2,2):&\quad -k_2\,q + (k_0+k_2)\,r = D_2, \tag{S15b}\\
(1,2)[\text{same as } (2,1)]:&\quad \Sigma\, q = k_1 r + k_2 p, \tag{S15c}
\end{align}

which reduce to Eqs.~(S2a)-(S2c), when $D_1=D_2=k_B T$. As in Sec.~S1, Eqs.~(S15a)-(S15b) give:

\begin{equation}
p = \frac{D_1+k_1q}{k_0+k_1}, \qquad r = \frac{D_2+k_2q}{k_0+k_2}. \tag{S16}
\end{equation}

Substituting Eq.~(S16) into Eq.~(S15c) and multiplying throughout by $(k_0+k_1)(k_0+k_2)$, one gets:

\begin{equation}
\Sigma q(k_0+k_1)(k_0+k_2) = k_1(k_0+k_1)D_2 + k_2(k_0+k_2)D_1 + k_1k_2\Sigma q,
\tag{S17}
\end{equation}

and again using $(k_0+k_1)(k_0+k_2)-k_1k_2=k_0\kappa$, Eq.~(S17) becomes:

\begin{equation}
\Sigma\kappa k_0\, q = k_2(k_0+k_2)D_1 + k_1(k_0+k_1)D_2,
\tag{S18}
\end{equation}

i.e.\ Eq.~(18) of the main text. Setting $D_1=D_2=k_B T$ in Eq.~(S18) reproduces Eq.~(S5), as expected.

\subsection*{S4.2 Antisymmetry of $QC$, for a general symmetric $D$}

The current is, $\bm J(\bm z)=Q\bm z\,P(\bm z)$, with $Q=DC^{-1}-A$ (replacing $k_B T\,\mathbf{I}\to D$ in Eq.~(S7), since the Fokker-Planck diffusion term is now $-D\nabla P$ rather than $-k_B T\,\mathbf 1\nabla P$). The antisymmetry argument of Sec.~S2 is also valid for \emph{any} symmetric $D$, using $CA^T=2D-AC$:

\begin{align}
QC &= D-AC, \notag\\
(QC)^T &= D - CA^T = D-(2D-AC) = -D+AC = -QC, \notag
\end{align}

so $QC+(QC)^T=0$ identically, exactly as before --- the isotropy of $D$ played no role in this step. Writing $QC=\tilde w\begin{pmatrix}0&-1\\1&0\end{pmatrix}$ and evaluating $(QC)_{21}=-(AC)_{21}=k_2p-(k_0+k_2)q$ using Eq.~(S16) for $p$, one obtains:

\begin{equation}
\tilde w = \frac{k_2 D_1 - q\,k_0\kappa}{k_0+k_1}. \tag{S19}
\end{equation}

Substituting $q$ from Eq.~(S18), $q\,k_0\kappa = [k_2(k_0+k_2)D_1+k_1(k_0+k_1)D_2]/\Sigma$, into Eq.~(S19), with further simplification yields:

\begin{equation}
\tilde w = \frac{k_2D_1\Sigma - k_2(k_0+k_2)D_1 - k_1(k_0+k_1)D_2}{\Sigma(k_0+k_1)}
= \frac{(k_0+k_1)\big[k_2D_1-k_1D_2\big]}{\Sigma(k_0+k_1)},
\tag{S20}
\end{equation}

using, $\Sigma-(k_0+k_2)=k_0+k_1$. Hence,

\begin{equation}
\tilde w = \frac{k_2D_1-k_1D_2}{\Sigma} = \frac{k_B(k_2T_1-k_1T_2)}{\Sigma} \equiv w,
\tag{S21}
\end{equation}

which is Eq.~(21) of the main text, and reduces to Eq.~(S8) when $T_1=T_2=T$. As in Sec.~S2, $Q=w\begin{pmatrix}0&-1\\1&0\end{pmatrix}C^{-1}$, giving the current components of Eq.~(20) of the main text with the same algebraic form as Eqs.~(S10a)-(S10b), only with $p,q,r,w$ now given by Eqs.~(S16),(S18),(S21).

The vanishing condition $w=0$, i.e. for $k_2T_1=k_1T_2$, generalizes both the known equilibrium limits: $k_1=k_2$ with $T_1=T_2$. However, $k_1=k_2$ with $T_1\neq T_2$ is \emph{not} equilibrium in general (as it reduces $w$ to $k_B k(T_1-T_2)/\Sigma$, as in case of a Brownian gyrator), while $k_1\neq k_2$ along with $T_1=T_2$ reduces to the non-reciprocal mechanism of the main text. The relation $k_1/k_2=T_1/T_2$ implies a scenario where these two independent symmetry-breaking mechanisms nullify the effects of one another.
\end{widetext}
\end{document}